# EPITAXIAL GROWTH OF CRYSTALLINE CaF$_2$ ON SILICENE

*Daniele Nazzari[†,*], Jakob A. Genser[†], Viktoria Ritter[†], Ole Bethge[§], Emmerich Bertagnolli[†], Tibor Grasser[#], Walter M. Weber[†], and Alois Lugstein[†,*]*

† Institute of Solid State Electronics, Technische Universität Wien, Gußhausstraße 25-25a, 1040 Vienna, Austria

§ Infineon Technologies Austria AG, Siemensstraße 2, 9500 Villach, Austria

# Institute for Microelectronics, Technische Universität Wien, Gußhausstraße 27-29, 1040 Vienna, Austria

E-mail: daniele.nazzari@tuwien.ac.at; alois.lugstein@tuwien.ac.at

ABSTRACT

Silicene is one of the most promising 2D materials for the realization of next-generation electronic devices, owing to its high carrier mobility, bandgap tunability and not least because of its intrinsic compatibility with mature Si processing technology. However, in order to take full advantage of the outstanding properties of silicene, it is necessary to engineer an insulating layer that can be interfaced directly to silicene without perturbing its bidimensional nature. At the same time, this insulating layer should exhibit low leakage currents even when highly scaled, to fully exploit the advantages of using a 2D material at the core of e.g a field effect device.

$CaF_2$ is known to form a quasi van der Waals interface with 2D materials, as well as to maintain its insulating properties even at ultrathin scales. Here we investigate the growth of thin $CaF_2$ layers on silicene by molecular beam epitaxy: diffraction images show that $CaF_2$ grows epitaxially on silicene/Ag(111), with its domains fully aligned to the 2D lattice. In-situ XPS analysis evidences that no changes in the chemical state of the silicon atoms can be detected upon $CaF_2$ deposition, excluding the formation of covalent bonds between Ca or F with Si. Polarized Raman analysis shows that silicene undergoes a structural change upon interaction with $CaF_2$, however retaining the bidimensional character and without transitioning to a $sp^3$-hybridized, bulk-like silicon.

INTRODUCTION

Epitaxial two-dimensional (2D) materials have attracted a high level of interest as they possess unprecedented optical and electronic properties and their synthesis process is easily scalable[1]. Among these, silicene is considered to be a particular attractive candidate[2,3] for the realization of next-generation high-performance devices, as it is characterized by an ultra-high carrier mobility[4] and a bandgap that can be tuned by the application of a perpendicular electric field[5,6]. Silicene is also expected to show intrinsic topological properties[7] and host Dirac cones at the K points of the Brillouin zone[8].

In order to engineer an e.g. silicene-based field effect device, it is necessary to develop a gating interface that couples with the 2D layer without altering its bidimensional nature and that is capable to withstand high-intensity electric fields.

In this regard, epitaxially grown calcium fluoride ($CaF_2$) is a promising material for the realization of such interface: it possesses a high dielectric constant (e=8.43) and a wide bandgap (E=12.1 eV), two characteristics that assure that tunnel leakages are negligible even for ultra-thin layers[9,10]. Most importantly, $CaF_2$ is fluorine terminated along the (111) direction, resulting in a completely inert surface[11,12], a prerequisite for preserving the properties of encapsulated 2D semiconductors. That's why it was also recently successfully employed for the realization of $MoS_2$-based field effect transistors with competitive device performances[10].

It has been shown that a high quality, single crystal $CaF_2(111)$ layer can be grown at relatively low temperatures of about 250°C on Si(111) by molecular beam epitaxy (MBE), thanks to the extremely small lattice mismatch (<1%) with the substrate. This moderate growth temperature is important to achieve pinhole-free films with better insulating properties[13], while remaining below the thermal stability limit of silicene[14].

In this study, we demonstrate the successful epitaxial growth of a thin layer of crystalline $CaF_2$ on 1 monolayer (ML) of silicene. Low energy electron diffraction (LEED) analysis shows that $CaF_2$ grows epitaxially on the 4x4, $\sqrt{13}$x$\sqrt{13}$ R13.9° and $2\sqrt{3}$x$2\sqrt{3}$ R30° phases of silicene on Ag(111), strictly following the orientation of the 2D silicon domains. Furthermore, by monitoring the Si2p core level by in situ X-ray photoelectron spectroscopy (XPS), we demonstrate that the Si atoms of the buried silicene layer are not forming covalent bonds with either Ca or F atoms. Finally, the vibrational properties of the buried silicene layer are analyzed by polarized Raman spectroscopy, evidencing a structural modification of the 2D layer. The encapsulated silicene retains its bidimensional nature. However, a clear shift of the main vibrational modes hints at an increased buckling and Si-Si bond length. In conclusion, here we demonstrate that crystalline $CaF_2$ can be epitaxially grown on silicene, paving the way for the realization of an ultra-thin gate insulation layer in silicene-based electronic devices.

METHODS

All the growth experiments were performed in an UHV-system with a base pressure of $5 \times 10^{-11}$ mbar. Pieces of Ag(111) on mica (MaTeck GmbH) were used as growth substrates and thoroughly cleaned in-situ through cycles of Ar+ sputtering and subsequent annealing at 520°C. Various phases of silicene were grown at substrate temperatures of 260°C or 300°C. The substrate temperature is measured using an infrared pyrometer (DIAS DGE10n) with a precision of +- 2 °C.

Silicene growth was achieved by Si evaporation from a rod (Goodfellow GmbH), using an electron beam evaporator (SPECS EBE-1), at a deposition rate of $\approx 0.02$ ML/min. For the actual growth of the insulating layer, $CaF_2$ crystals (Sigma-Aldrich, GmbH) were evaporated from a tungsten crucible mounted in a second electron beam evaporator (SPECS EBE-4) at a rate of $\approx 0.02$ nm/min.

LEED analysis was performed using the SPECS ErLEED 100 optics at an electron energy of 35 eV and at low exposure times to prevent damages to the $CaF_2$ layer. XPS and UPS (UV photoelectron spectroscopy) analysis were also performed in-situ. Photons with an energy of 1486.65 eV are generated using a SPECS XR50 source equipped with an Al anode for XPS analysis. For UPS, photons with an energy of 21.2 eV are generated by a He plasma discharge using a SPECS UVS 10 source. The emitted photoelectrons are collected at an angle of 60° in a SPECS 150 hemispherical analyzer and detected by a 2D CCD detector. The acquired spectra are analyzed using the software CasaXPS: firstly, the background is subtracted recurring to a U 2 Tougaard model; peaks are then fitted using Voigt functions.

Finally, ex-situ polarized Raman spectroscopy is performed on the sample immediately after removal from the UHV system, to prevent oxidation of the silicene layer. The analysis is performed in a back-scattering geometry using a confocal $\mu$-Raman setup (Alpha 300, WITec) equipped with a 532 nm Nd:YAG laser.

DISCUSSION

Fig. 1a shows the LEED pattern of 1 ML of silicene grown on Ag(111) at a growth temperature of 260°C. The diffraction pattern can be well interpreted assuming two silicene phases that are expected for the lower growth temperature[15–17], as shown by the superimposed diffraction model in the right hemisphere. The red pattern is related to the formation of a 4x4 silicene phase, while the blue signal reflects the diffraction pattern related to the $\sqrt{13}$x$\sqrt{13}$R13.9° one. In both cases, the Wood notation describes the size of the supercells with respect to the Ag(111) crystal and their orientation relative to the Ag lattice. In all the images, the white circles represent the position of the first order diffraction spots of Ag(111).

Overall, these supercells account for a total of 5 different orientations of the silicene layer on top of the Ag(111) substrate. Specifically, the 4x4 one accounts for a structure where the silicene[10] vector is aligned along Ag[10], while the $\sqrt{13}$x$\sqrt{13}$R13.9 supercell comprises 4 different silicene domains, where the angle between the two vectors is either $\pm 5.2°$ or $\pm 33°$, as clearly demonstrated by Resta et al.[17]

After having acquired the LEED pattern, the $CaF_2$ evaporation is ramped up and ~ 1 nm of $CaF_2$ is deposited, at the same growth temperature of 260°C.

The relatively low sample temperature for the $CaF_2$ growth was chosen for two reasons: firstly, it has been reported that $CaF_2$ layers grown at this temperature are pinhole-free[18]; secondly, the growth temperature must be low enough to preserve the silicene layer, which is known to degrade for temperatures higher than ~ 325°C.[19]

The LEED pattern in Fig. 1b, recorded after the $CaF_2$ deposition, is drastically different compared to the one of silicene, shown in Fig. 1a, with a markedly reduced number of diffraction spots.

This pattern, however, can be easily modelled by assuming the presence of different $CaF_2$ domains, with several orientations with respect to Ag(111). As shown in the right half of Fig. 1b, it is therefore possible to clearly identify domains with a rotation of 0° (yellow), ± 5.2° (purple) and ± 33° (green) with respect to Ag[10]. These orientations are exactly the same ones observed for the silicene layers[17], suggesting that the growth of the $CaF_2$ domains is precisely aligned with the underlying silicene phases.

If the same quantity of $CaF_2$ is deposited directly on a bare Ag(111) substrate, the observed diffraction pattern is drastically different, as shown in Fig. 1c. In this case, the diffraction pattern is composed by only 6 diffraction spots associated to $CaF_2$ domains aligned with the Ag(111) ones and a bright diffraction ring, determined by the presence of many randomly oriented $CaF_2$ domains, likely due to the large lattice mismatch between $CaF_2$ and Ag(111).

It is well documented that, by increasing the growth temperature of silicene to 300°C, it is possible to obtain an additional silicene phase, described by supercell 2√3x2√3R30°, as shown in Fig. 1d, corresponding to a 2D lattice rotated by ~ 10.9° with respect to Ag(111).[15,20]

As shown in the right hemisphere of Fig. 1d, the diffraction spots can be well modelled by taking into account the high temperature silicene phase (light blue), alongside the already mentioned 4x4 phase (red). Fig. 1e shows the diffraction pattern after the growth of 1 nm thick $CaF_2$ layer at a temperature of 260°C. The diffraction pattern can be well modelled by considering $CaF_2$ domains with an angle of 0° (yellow) and 10.9° (pink) with respect to the Ag[10] direction, matching again perfectly the orientation of the underneath silicene phases.

It is thus clear that the presence of just a single layer of silicene is able to direct the growth of $CaF_2$. The absence of silicene-specific spots does not mean that the crystalline structure of silicene is lost, as LEED cannot probe the buried 2D layer, a direct consequence of the extremely small inelastic mean free path of the electrons employed in this technique[21,22]. It rather indicates that the 1 nm, i.e. approx. 2 ML thick, $CaF_2$ layer completely covers the silicene layer.

Such a precise correspondence between the $CaF_2$ domains and the silicene layer might be a sign of the formation of covalent bonds between Si and Ca or F atoms or it may be driven by the energetically favorable alignment of their respective lattices due to the extremely small lattice mismatch between silicene and $CaF_2$.

To obtain more precise information on the silicene/$CaF_2$ interface properties, the sample is analyzed using XPS, before and after $CaF_2$ deposition. From now on, all investigations will refer to the silicene layer obtained at 260°C, as the high temperature phase is known to be highly defective[23].

Fig. 2 shows the XPS signal of the Si2p peak, indicative of the chemical state of the Si atoms in the 2D layer, before (top) and after (bottom) the $CaF_2$ deposition onto the silicene/Ag stack. The spectra in the main panels show the Si 2p doublet (green) and the Ag 4s singlet (blue) peaks. Spectra are collected at a takeoff angle of 60°, to improve the sensitivity towards silicene and to reduce the intensity of the Ag4s, which partially overlaps with the Si2p peak. The Si2p peak, accounting for the Si2p$_{1/2}$ and the Si2p$_{3/2}$ components, shifts by only 31 meV after the CaF$_2$ deposition, likely due to a charge transfer between the CaF$_2$ layer and the silicene. It is known that fluorine atoms induce a positive charge on the Si atoms, causing a ~ 0.9 eV shift of the Si2p peak towards higher binding energies upon formation of Si-F bonds[24,25], while Ca atoms, being less electronegative than Si, lead to a shift of ~ 0.45 eV in the opposite direction[24,26].

Having observed such a small shift of the Si2p peak upon CaF$_2$ deposition, the formation of covalent bonds between Si and Ca or F can be excluded. In contrast, pronounced shifts of the Si2p peak were observed in previous experiments where CaF$_2$ was grown on regular, sp$^3$-hybridized Si and at higher temperatures, strongly suggesting the formation of Ca-Si and F-Si bonds at the interface[24,27,28]. The insets show the respective XPS spectra at the energy ranges relative to the F1s and the Ca2p peaks. After CaF$_2$ deposition, the characteristic Ca2p and F1s peaks can be observed with a corresponding stoichiometry of 1:2.

To probe the valence band properties of the deposited CaF$_2$, UV photoemission spectroscopy is used. The collected spectrum, shown in Fig. 3, is characterized by a narrow peak at 16.58 eV signaling the onset of secondary electrons emission and by a wider peak centered around 10 eV, compatible with the valence states of CaF$_2$[28]. The peak found at lower binding energies can be well fitted using a 2 components model, as shown in the inset. This double-peaked nature of the valence band is typical for bulk alkali halides and it is predicted by band-structure calculations[29]. The fitted peaks are centered at 9.3 eV and at 10.4 eV, in line with previous results[28], showing that the deposited film is in pristine conditions.

In contrast to the actual CaF$_2$ deposition on silicene at 260°C, the much higher deposition temperatures (500°C-750°C) used for the investigations on Si(111) mentioned above, most probably leads to a decomposition of $CaF_2$ at the interface and therefore the formation of distinct Si-F and Ca-Si bonds.

Furthermore, one should keep in mind that the Si2p signal of Fig. 2 is generated by a bidimensional layer of Si, therefore it is related exclusively to atoms that are interfacing $CaF_2$ without contributions related to bulk material. Since $CaF_2$ is completely covering the silicene layer, as deduced from LEED images, all Si atoms are interfacing the $CaF_2$ fluoride layer. This means that the formation of a chemical bond different from the existing Si-Si would induce a complete shift in the Si2p peak and not just a modification of its shape, a condition clearly ruled out by the collected data.

The current XPS data thus clearly demonstrate that the chemical state of the Si atoms remains unchanged upon $CaF_2$ deposition, but do not allow any statements to be made regarding the structure of the silicene layer.

In order to investigate possible structural modifications of silicene, a thicker $CaF_2$ layer (10 nm) was deposited at the same growth temperature, acting as an effective passivation layer and thus enabling ex-situ Raman investigations.

The polarized Raman spectra for the $CaF_2$ covered silicene layer, are shown in Fig. 4, for parallel (red) and crossed (blue) polarization configurations where the vectors of the incident and scattered light are, respectively, parallel and normal to each other.

Two main Raman peaks are located at 388 cm$^{-1}$ and 457 cm$^{-1}$. The observed Raman spectrum deviates strongly from the ones observed from a bare silicene layer on Ag(111), with the characteristic Raman peaks located at 175 cm$^{-1}$, 216 cm$^{-1}$ and 514 cm$^{-1}$ [14,30].

This drastic change in the Raman spectrum is a clear signal of structural modifications of silicene due to the presence of the $CaF_2$ encapsulation. This is obviously different to what we have shown for silicene when interfaced with a van der Waals material, like graphene, where the extremely low out-of-plane forces are not able to change the structure of the 2D material, leaving the Raman signature unaltered[30]. In contrast it was shown that, upon intentional hydrogenation, silicene undergoes a rehybridization from a mixed $sp^2$-$sp^3$ character to a dominant $sp^3$ one, with an increased buckling of the Si atoms[31].

When analyzing the structural changes due to the $CaF_2$ deposition, it is important to firstly note that these do not lead to the formation of bulk-like, $sp^3$-hybridized Si clusters, as these would generate a sharp and very intense peak located at 520 cm$^{-1}$ [30], which obviously cannot be observed in the spectrum shown in the lower panel of Fig. 4. When the Raman spectrum of the $CaF_2$ capped silicene is collected in a crossed polarization configuration (blue spectrum), the intensity of the vibrational mode located at 388 cm$^{-1}$ is clearly reduced, while the one of the 457 cm$^{-1}$ peak is similar in both configurations, demonstrating that the two observed peaks must be related to phonon modes with different symmetries. Notably, the peak located at 388 cm$^{-1}$ is also observed in $CaSi_2$ structures, related to the out-of-plane vibrations of the Si atoms, that are arranged in 2D, silicene-like planes stacked between Ca planes[32]. The observed polarization dependency and the similarity to the $CaSi_2$ spectrum strongly suggest that the 388 cm$^{-1}$ mode is related to the out of plane vibration of the Si atoms, with a strong blueshift compared to bare silicene on Ag(111). It is then reasonable to assign the 457 cm$^{-1}$ vibrational mode to the in-plane vibration of Si atoms: this mode is therefore strongly redshifted if compared to the one detected in silicene/Ag(111)[14,30], but blueshifted with respect to the in-plane mode of Si atoms in $CaSi_2$ structures[32].

It is important to note that, differently from the case of bare silicene on Ag(111), the out-of-plane vibrational mode is not totally suppressed in cross polarization configuration. This is a clear indication that, due to structural modifications, the silicene layer is now less symmetric, not belonging anymore to group C6v. This deformation, however, does not lead to the formation of bulk-like sp³ hybridized Si and comes without the formation of covalent bonds between Si and either F or Ca atoms, as proven by the XPS measurements. The detected changes in the silicene Raman spectrum point to an increase of the in-plane Si-Si bond length, resulting in a redshift of the in-plane vibrational mode[33].

Lengthening the Si-Si bonds while maintaining the basic structure leads to increased buckling, which would also explain the blueshift of the out-of-plane Raman peak, similarly to what is observed upon hydrogenation of silicene/Ag(111)[31].

CONCLUSIONS

To summarize, we have investigated $CaF_2$ on top of a stack composed by 1 ML of silicene grown on Ag(111) as a promising material for the realization of an effective gate insulation layer in silicene-based electronic devices. LEED patterns prove that the $CaF_2$ layer is crystalline and grows following the orientation of the various silicene phases that form on Ag(111). XPS analysis shows unambiguously that the Si2p peak of the silicene layer does not shift upon deposition of $CaF_2$, which excludes the formation of covalent bonds between Ca, F and Si atoms. Polarization dependent Raman spectroscopy of the buried layer evidences that silicene undergoes structural modifications when interfaced with $CaF_2$, however maintaining the 2D nature. The shifts of the main Raman components are indicative of an increased Si-Si bond length and of a larger separation between the two buckled planes of Si atoms. Our results demonstrate that silicene can be successfully interfaced with $CaF_2$, without losing its bidimensional character. It is important to consider that the structural modification that silicene exhibits may lead to a significant change in the electronic transport properties, which should be carefully assessed in a follow-up study.

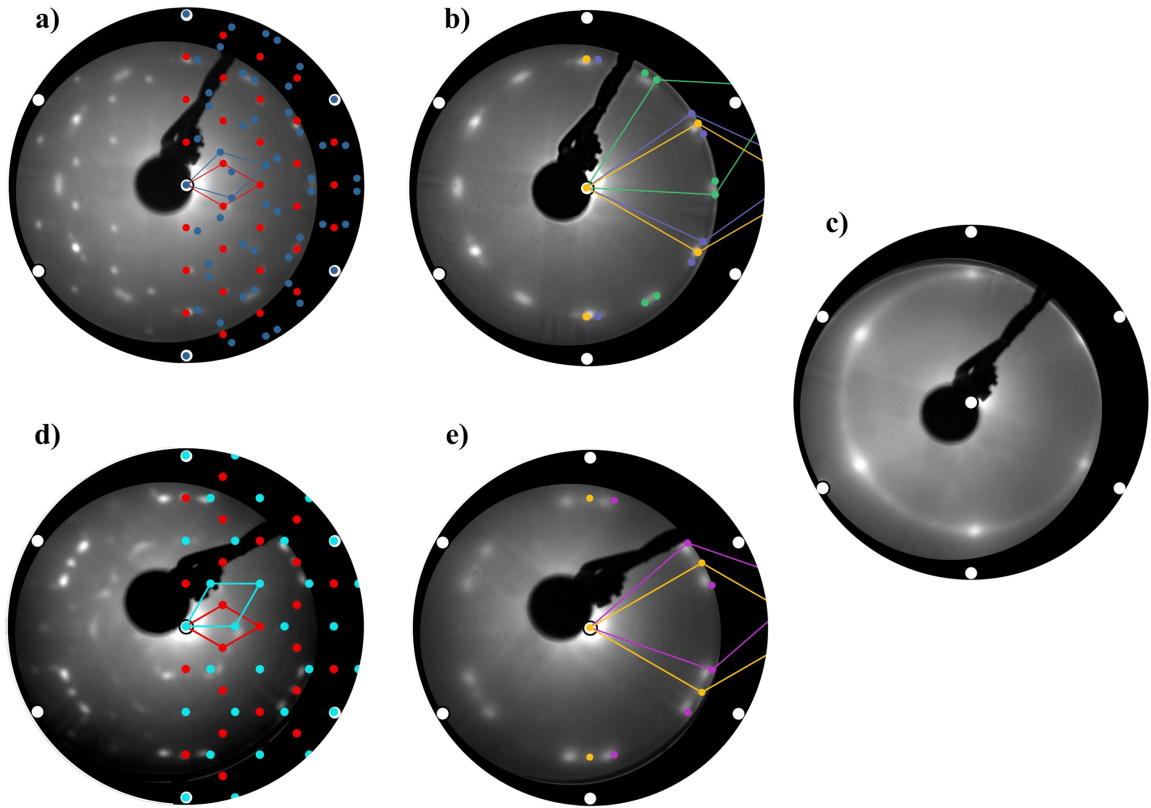

Fig 1. LEED patterns acquired at an energy of the primary electron beam of 35 eV. In all the images, the white disks represent the position of the first order diffraction spots of Ag(111). The shown patterns correspond to: a) 1 ML of silicene grown on Ag(111) at 260°C, composed of a 4x4 phase (red) and of a √13x√13R13.9 one (blue); b) a thin layer of CaF$_2$ deposited ontop of the silicene layer shown in a). Different CaF$_2$ domains are visible, oriented with an angle of 0° (yellow), +-5.2° (purple), +-33° (green), with respect to the Ag[10] direction; c) a thin layer of CaF$_2$ deposited on Ag(111); d) 1 ML of silicene grown on Ag(111) at 300°C. The pattern is composed of a 4x4 phase (red) and of a 2√3x2√3R30 one (light blue). e) a thin layer of CaF$_2$ deposited on the same silicene layer shown in d). CaF$_2$ domains are oriented with an angle of 0° (yellow) and 10.9° (pink), with respect to the Ag[10] direction.

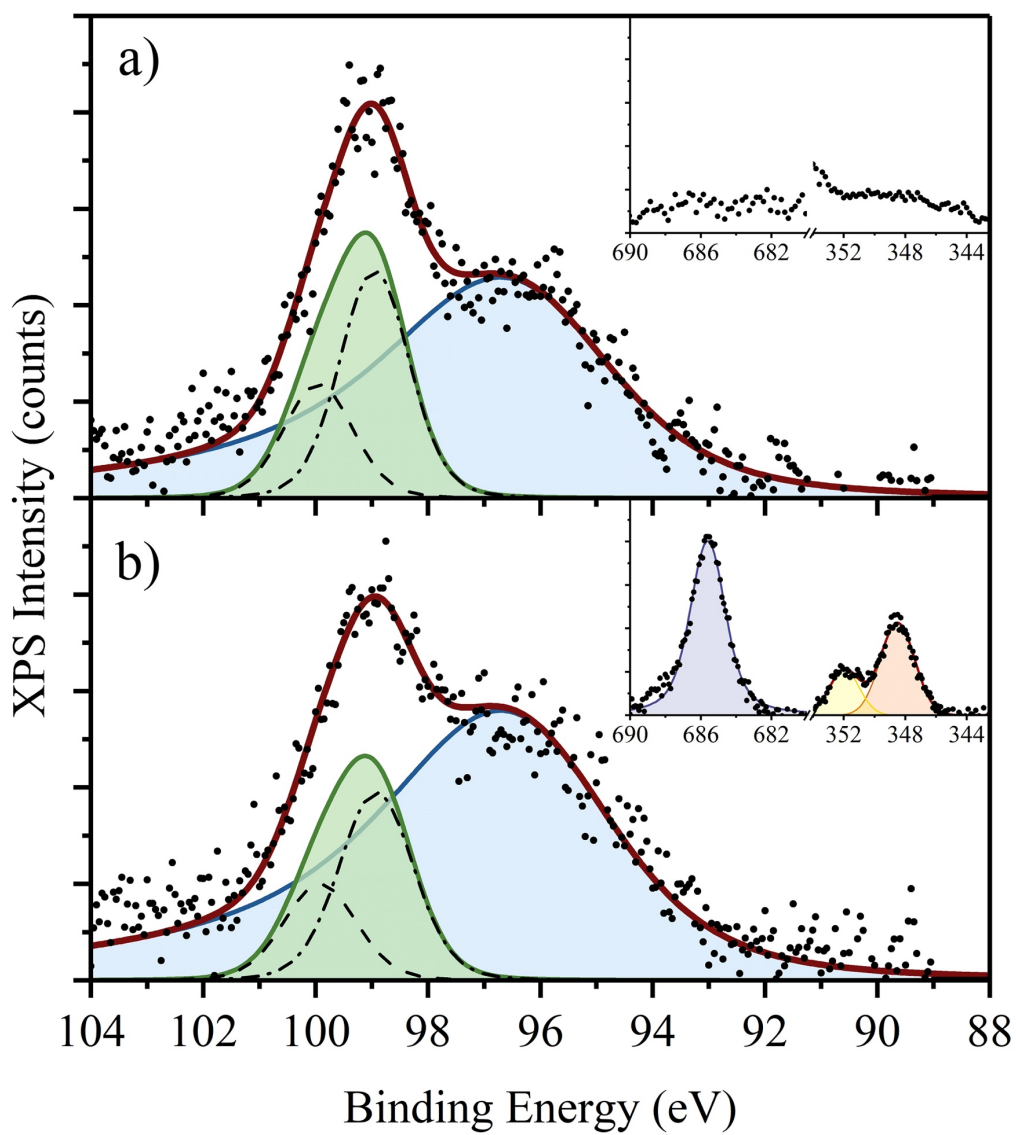

Fig 2. a) XPS spectrum of the Si2p peak of 1 ML silicene grown on Ag(111) at 260°C. The spectrum is composed by the Si 2p doublet (green) and the Ag 4s singlet (blue). In the inset a scan of the F1s and Ca2p regions is presented. b) XPS spectrum of the Si2p peak analyzed after the deposition of the $CaF_2$ layer. A scan of the F1s and Ca2p regions confirms the presence of CaF2, as shown in the inset.

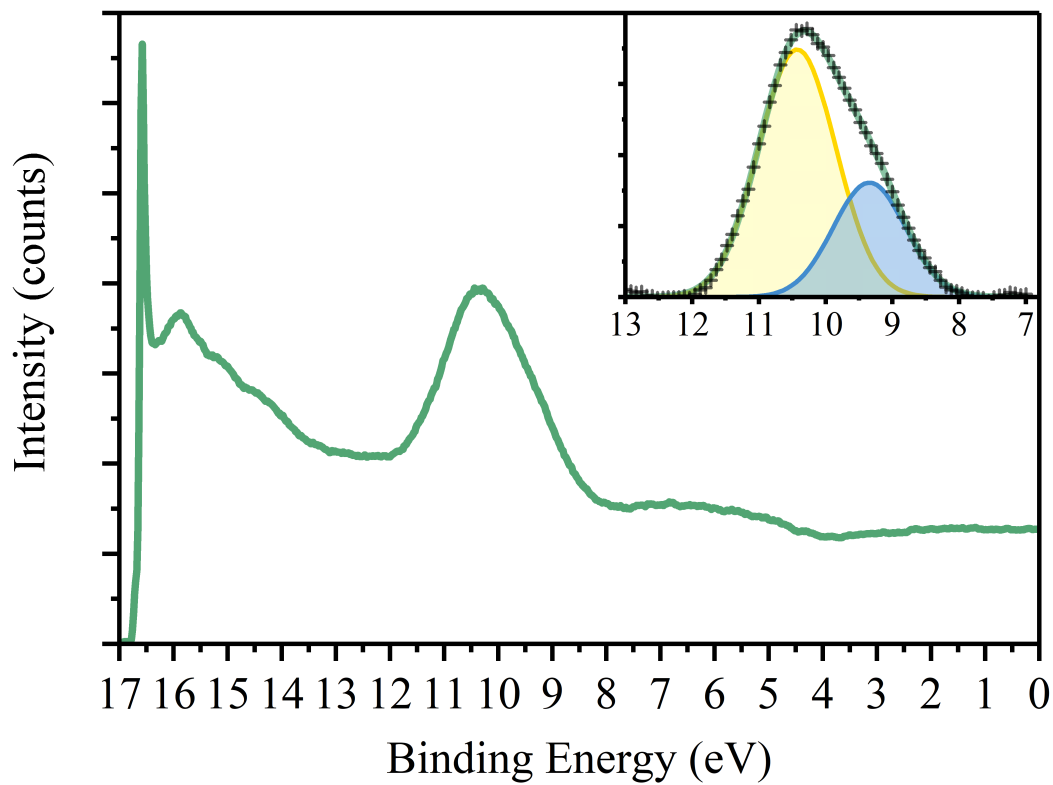

Fig 3. UPS spectrum of 1 nm $CaF_2$ deposited on 1 ML of silicene/Ag(111). The sharp peak at 16.58 eV denotes the onset of secondary electrons emission. The broad peak centered at ~10 eV is related to the valence states of $CaF_2$. In the inset, the fit shows that the broad peak can be modelled by considering a two-components model.

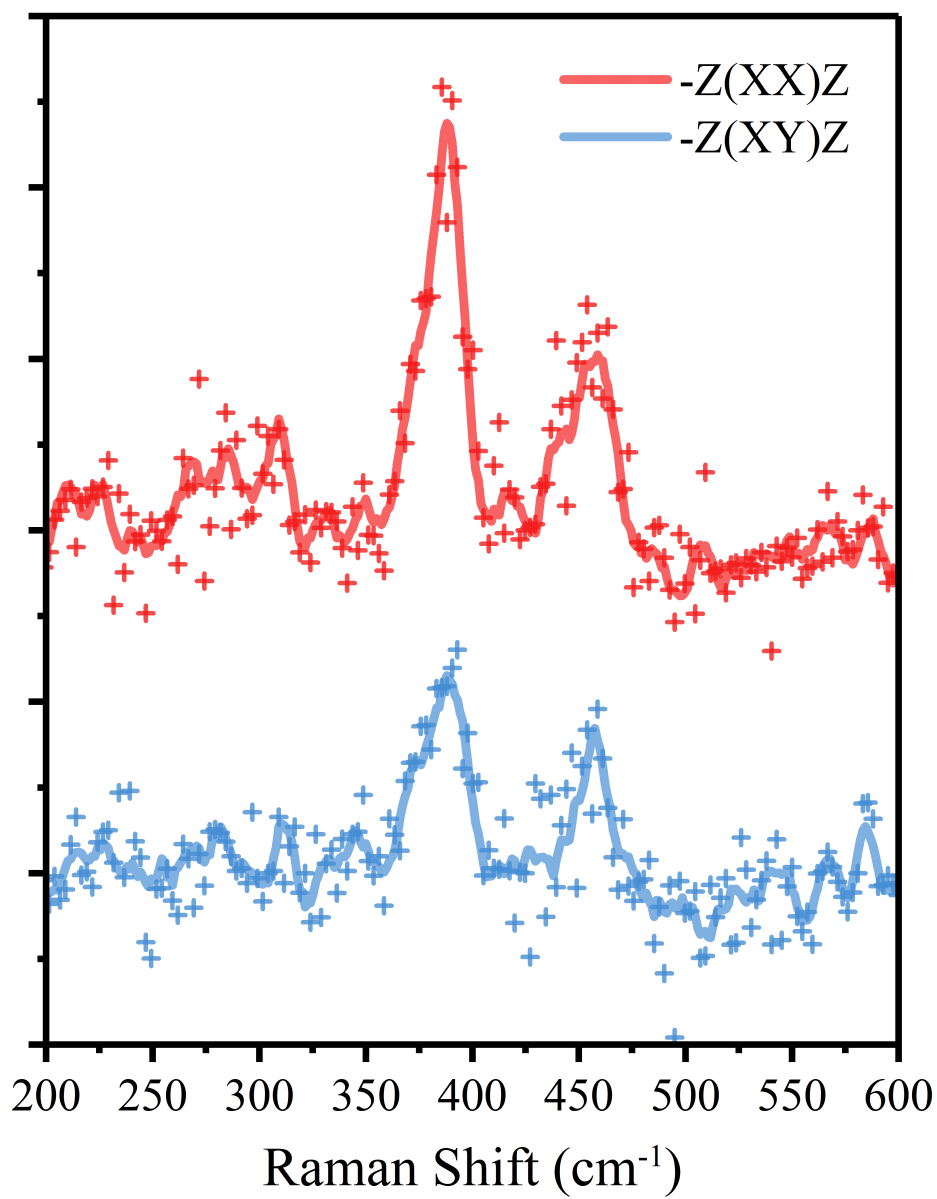

Fig 4. Ex-situ Raman spectrum of 1 ML silicene grown on Ag(111) at 260 °C and covered by 10 nm $CaF_2$, collected in parallel (red) and crossed (blue) polarization configurations.

ACKNOWLEDGMENT

This work was funded by the Fonds zur Förderung der Wissenschaftlichen Forschung (FWF), Austria (project P29244-N27). We thank the center for micro- and nanostructures (ZMNS) at the TU Vienna for access to the cleanroom facilities. We would like to thank Prof. Nikolai S. Sokolov, Dr. Yury Yu. Illarionov for the support in the evaporation of $CaF_2$.